\documentclass{article}

\usepackage{spconf,amsmath,amsfonts,graphicx}
\usepackage{amssymb,textcomp,mathtools}
\usepackage{bm,upgreek,algorithm,hyperref}
\usepackage{multirow,booktabs,hhline,array}
\usepackage{cite,url,makecell,setspace}

\pdfoutput=1

\def\thline{\noalign{\hrule height 1.0pt}}

\renewcommand{\vec}[1]{\bm{\mathrm{#1}}}

\title{Implicit Filter-and-sum Network for Multi-channel Speech Separation}
\name{Yi~Luo, Nima~Mesgarani}
\address{
  Department of Electrical Engineering, Columbia University
}

\begin{document}
\ninept
\maketitle

\begin{abstract}
Various neural network architectures have been proposed in recent years for the task of multi-channel speech separation. Among them, the filter-and-sum network (FaSNet) performs end-to-end time-domain filter-and-sum beamforming and has shown effective in both ad-hoc and fixed microphone array geometries. In this paper, we investigate multiple ways to improve the performance of FaSNet. From the problem formulation perspective, we change the explicit time-domain filter-and-sum operation which involves all the microphones into an implicit filter-and-sum operation in the latent space of only the reference microphone. The filter-and-sum operation is applied on a context around the frame to be separated. This allows the problem formulation to better match the objective of end-to-end separation. From the feature extraction perspective, we modify the calculation of sample-level normalized cross correlation (NCC) features into feature-level NCC (fNCC) features. This makes the model better matches the implicit filter-and-sum formulation. Experiment results on both ad-hoc and fixed microphone array geometries show that the proposed modification to the FaSNet, which we refer to as iFaSNet, is able to significantly outperform the benchmark FaSNet across all conditions with an on par model complexity.
\end{abstract}

\section{Introduction}
\label{sec:introduction}
The design of multi-channel speech separation systems is one of the active topics in the speech separation community in the past years. Despite the advances in time-frequency domain neural beamformers where a neural network is used to assist the conventional beamformers for better robustness and performance \cite{heymann2016neural, erdogan2016improved, xiao2017time, ochiai2017unified, zhang2017speech, heymann2018performance, qian2018deep, xu2019joint, ochiai2020beam, xu2020neural, zhang2020adl}, time-domain architectures have also earned the attention from the community due to their ability to perform purely end-to-end optimization towards the target signals. Moreover, as conventional beamformers often require a temporal context for better estimation of spatial features \cite{higuchi2017online, higuchi2018frame}, time-domain systems have the potential to be operated at frame-level with a lower theoretical system latency.

Recent time-domain systems can be classifies into three categories. The first category reformulates the multi-channel separation problem as a single-channel separation problem on a selected reference microphone, with the help of additional cross-channel features. Various cross-channel features have been proposed and analyzed in versatile datasets \cite{gu2019end, gu2020temporal, gu2020enhancing, koyama2020efficient, fan2020spatial}. The second category processes the multi-channel mixtures with a convolutional encoder where the channels are treated as different feature maps in a convolutional operation. Such systems learn a direct mapping between the mixtures and the target signals \cite{stoller2018wave, liu2020multichannel}.  The third category performs end-to-end beamforming without solving the optimization problems required in conventional beamformers. The beamforming filters can be compared to the masks in single-channel separation systems as they both operate at frame-level \cite{luo2019fasnet, luo2020end}.

One of the systems in the third category is the filter-and-sum network (FaSNet) which performs time-domain end-to-end beamforming \cite{luo2019fasnet, luo2020end}. FaSNet attempts to directly estimate the beamforming filters via a neural network, and previous results have proven its effectiveness in simulated noisy reverberation datasets comparing with single-channel methods and baseline methods in the first category. However, FaSNet has several drawbacks in the problem formulation. Since FaSNet formulates the separation problem as a time-domain beamforming problem and the end-to-end training target is typically the reverberant clean signals, asking the filters to reconstruct not only the direct-path signal but also all the reverberation components may not be optimal, as it requires the filters to achieve a complex beampattern to preserve the reverberations. Although it is possible to perform joint dereverberation and separation by setting the direct-path signals as the training target, it may significantly increase the task difficulty. Moreover, one reason for the preference of time-frequency domain beamformers than time-domain beamformers is their advantage in both robustness and performance due to the use of short-time Fourier transform (STFT), hence it is also necessary to explore such beamforming operation in a latent space.

In this paper, we explore multiple aspects to improve the original FaSNet architecture. We investigate four modifications to FaSNet. First, we compare the original multi-input-multi-output (MIMO) formulation with the multi-input-single-output (MISO) formulation, where the filter is only estimated for the reference channel instead of all the channels. Second, we consider estimating the filter in a learnable latent space instead of the original waveform-domain. Together with the MISO assumption, it also matches the design of the systems in the first category. Third, we look for better cross-channel features more suitable for the MISO and latent-space filtering design. Fourth, we utilize context-aware processing to further improve the model performance. We refer to the modification of FaSNet as implicit FaSNet (iFaSNet). Ablation experiments show that such modifications are able to allow iFaSNet to outperform the original FaSNet across various data configurations while maintaining a same model size and complexity.

The rest of the paper is organized as follows. Section~\ref{sec:fasnet} briefly goes over the original FaSNet architecture and introduces the proposed iFaSNet. Section~\ref{sec:config} provides the experiment configurations. Section~\ref{sec:result} presents the results and discussions. Section~\ref{sec:conclusion} concludes the paper.

\section{Implicit Filter-and-sum Network}
\label{sec:fasnet}
\begin{figure*}[!ht]
	\small
	\centering
	\includegraphics[width=2\columnwidth]{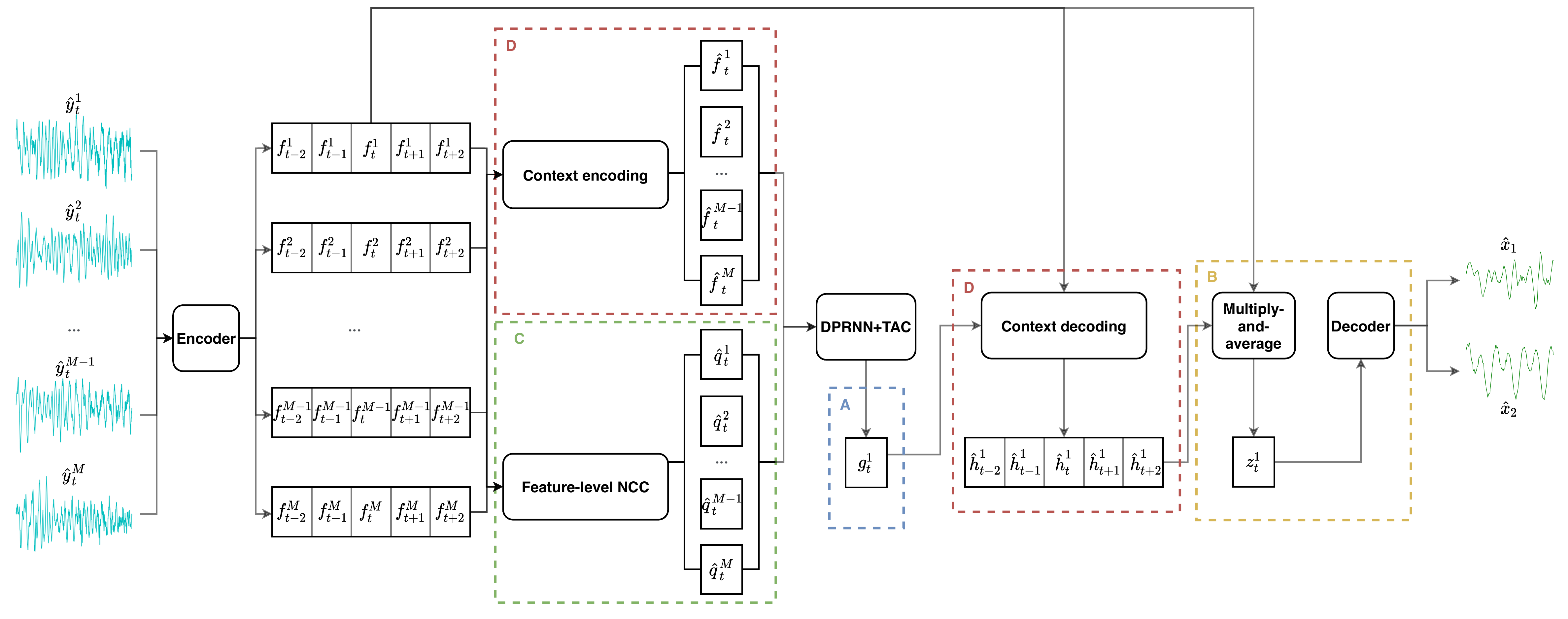}
	\caption{Flowchart for the proposed iFaSNet architecture. The modifications to the original FaSNet are highlighted, which include (A) the use of MISO design instead of the original MIMO design, (B) the use of implicit filtering in the latent space instead of the original explicit filtering on the waveforms, (C) the use of feature-level NCC feature for cross-channel information instead of the original sample-level NCC feature, and (D) the use of context-aware filtering instead of the original context-independent filtering.}
	\label{fig:flowchart}
\end{figure*}

\subsection{Filter-and-sum Network recap}
\label{sec:FaSNet}

Filter-and-sum network (FaSNet) performs time-domain filter-and-sum beamforming at frame level and directly estimates the beamforming filters with a neural network. For each frame of input mixtures from the $M$ channels $\{\vec{y}^i\}_{i=1}^{M} \in \mathbb{R}^{1\times L}$, a context window of $W$ samples in both past and future is concatenated into $\vec{y}^i$, resulting in a context frame $\hat{\vec{y}}^i \in \mathbb{R}^{1\times(L+2W)}$. For each target source, $M$ time-domain beamforming filters $\{\vec{h}^i\}_{i=1}^{M} \in \mathbb{R}^{1\times (1+2W)}$ are estimated from $\{\vec{y}^i\}_{i=1}^{M}$ by a neural network, and the filters are applied to the input to obtain the separated outputs:
\begin{align}
	\hat{\vec{x}} = \sum_{i=1}^M \hat{\vec{y}}^i \circledast \vec{h}^i
\label{eqn:fas}
\end{align}
where $\circledast$ represents the convolution operation. The estimation of the filters rely on both the channel-wise features and the cross-channel features, and FaSNet applies a linear fully-connected (FC) layer to extract the channel-wise features for each input channel:
\begin{align}
    \vec{s}^i = \hat{\vec{y}}^i\hat{\vec{W}}
\label{eqn:enc}
\end{align}
where $\hat{\vec{W}} \in \mathbb{R}^{(L+2W)\times N}$ is a learnable parameter matrix. The linear FC layer can be also regarded as a linear 1-D convolutional layer. The cross-channel feature used by FaSNet is the normalized cross correlation (NCC) feature $\vec{q}^i \in \mathbb{R}^{1\times (1+2W)}$ calculated between the center frame at reference microphone $\vec{y}^1$ and the context frame at all microphones $\{\hat{\vec{y}}^i\}_{i=1}^{M}$:
\begin{align}
\begin{cases}
    \hat{\vec{y}}_j^i = \hat{\vec{y}}^i[j:j+L-1] \\
	q_j^i = \frac{\vec{y}_1\hat{\vec{y}}_j^{iT}}{\left\|\vec{y}_1\right\|_2 \left\|\hat{\vec{y}}_j^i\right\|_2}
\end{cases}, \quad j=1, \ldots, 2W+1
\label{eqn:tNCC}
\end{align}

Various neural network architectures can be selected to estimate the filters. FaSNet makes use of the dual-path RNN (DPRNN) \cite{luo2020dual} architecture together with the transform-average-concatenate (TAC) module \cite{luo2020end} to perform robust filter estimation and allows the model to be invariant to the number and permutation (locations) of the microphones. This is rather important in ad-hoc array scenarios.

\subsection{Implicit Filter-and-sum Network}
\label{sec:iFaSNet}

To improve the performance of FaSNet, we propose the implicit filter-and-sum network (iFaSNet) which modifies the standard FaSNet in multiple aspects. Figure~\ref{fig:flowchart} shows the flowchart of iFaSNet, and the modifications to the original FaSNet are highlighted.

\subsubsection{Multi-input-single-output design}
\label{sec:MISO}

FaSNet applies the time-domain filter-and-sum beamforming together with an end-to-end training objective. The training target for FaSNet is typically set to the reverberant clean signals in a selected reference microphone\footnote{We do not consider the case for joint separation and dereverberation here.}. However, the learned beamforming filters at different channels have their own beampatterns, and using the reverberant clean signals as the training target implies that the filters should not only enhance the signal coming from a certain direction, but also need to reconstruct all the reverberation components. The corresponding beampattern can thus be a mess, as the reverberations typically cover a much wider range of angles. Although FaSNet applies frame-level beamforming where infinite optimal frame-level filters may exist since the linear equation in equation~\ref{eqn:fas} is underdetermined ($M\times (1+2W)$ unknowns and $1+2W$ equations), finding such reverberation-preserving filters for all channels may hurt the generalization of the network and thus affect the separation performance.

To maintain the end-to-end training configuration, we change the original multi-input-multi-output (MIMO) filter estimation into multi-input-single-output (MISO) estimation, where only the filter for the reference channel is calculated. This reformulates the multi-channel separation problem back to the single-channel separation problem as in the first category discusse in Section~\ref{sec:introduction}, while the input to the model still contains the mixtures from all channels. Figure~\ref{fig:flowchart} (A) shows the MISO module.

\subsubsection{Implicit filtering in latent space}
\label{sec:implicit}

FaSNet explicitly calculates the beamforming filters in the waveform-domain. The advantage is that it follows the standard definition of time-domain beamforming, and it is easy to analyze the filters' behaviors such as beampatterns. However, most literatures on multi-channel separation focus on learning a mapping in a latent space where the signal can be better represented. Not only the existing neural beamformers are mainly designed in the time-frequency domain, but also the time-domain systems utilize a learnable latent space for better signal representations and separation performance. Motivated by these recent progress, we adopt the standard setting in time-domain systems where a pair of learnable encoder and decoder are used for signal representation in the latent space. The filter is thus similar to the ``mask'' in single-channel separation systems, which is defined as a multiplicative function on the encoder output of the reference channel.

According to equation~\ref{eqn:enc}, the encoder in the original FaSNet maps the context frame $\vec{\hat{y}}^i$ to a feature vector $\vec{s}^i$ by the encoder weight $\vec{W}$. $\vec{s}^i$ is only used for the estimation of the beamforming filters and does not involve in the beamforming operation. The encoder in iFaSNet is applied on the center frame $\vec{y}^i$ instead of the context frame $\hat{\vec{y}}^i$, with its corresponding encoder weight $\vec{W} \in \mathbb{R}^{L\times N}$. The feature is denoted as $\vec{f}^i \in \mathbb{R}^{1\times N}$. A decoder with its weight $\vec{U} \in \mathbb{R}^{N\times L}$ is also applied to transform the latent feature back to waveforms. Figure~\ref{fig:flowchart} (B) shows the newly-added decoder module.

\subsubsection{Feature-level normalized cross correlation}
\label{sec:fNCC}

The cross-channel feature in original FaSNet is calculated by time-domain normalized cross correlation (tNCC) defined in equation~\ref{eqn:tNCC}. The rationale behind tNCC is to capture both the delay information across channels and the source-dependent information for different targets. However, the tNCC feature requires a sample-level convolution operation, which involves $L(1+2W)M$ float-point multiplications. Moreover, when the sample rate of the signals is low (e.g. in certain telecommunication systems), such sample-level correlation may fail to capture the cross-channel delay information. To save the number of operations and accelerate the feature calculation, we modify the tNCC to feature-level NCC (fNCC). Denote the context feature $[\vec{f}^i_{t-C}, \ldots, \vec{f}^i_t, \ldots, \vec{f}^i_{t+C}]$ as $\vec{F}^i_t \in \mathbb{R}^{(1+2C)\times N}$ where $C$ denotes the context size. fNCC calculates the cosine similarity between $\vec{F}^1_t$ and $\{\vec{F}^i_t\}_{i=1}^M$:
\begin{align}
    \hat{\vec{q}}^i_t = \bar{\vec{F}}_t^1\bar{\vec{F}}^{iT}_t
\label{eqn:fNCC}
\end{align}
where $\bar{\vec{F}}^i_t$ denotes the column-normalized feature of $\vec{F}^i_t$ where each column has a unit length, and $\hat{\vec{q}}^i_t \in \mathbb{R}^{(1+2C)\times (1+2C)}$ denotes the fNCC feature at time $t$ for channel $i$. $\hat{\vec{q}}^i_t$ is then flatten to a vector of shape $1\times (1+2C)^2$.

fNCC only contains $N(1+2C)^2M$ float-point multiplication operations. For the default setting in FaSNet where $W=L=256$ with a 50\% overlap between frames, we have $C=2$ and $(1+2C)^2=25 \ll 1+2W=513$. By properly setting the value of $N$, fNCC can greatly save the computation needed for the cross-channel feature extraction. Figure~\ref{fig:flowchart} (C) shows fNCC calculation module.

\subsubsection{Context-aware filtering}
\label{sec:context}

Existing systems for both single- and multi-channel end-to-end separation only make use of the center feature $\vec{f}^i_t$ at time $t$. On the other hand, utilizing context information to improve the modeling of local frame is very common in various systems \cite{xu2014regression, yu2017permutation}. iFaSNet explores a context encoder to squeeze the context feature $\vec{F}^i_t$ into a single feature vector $\hat{\vec{f}}^i_t \in \mathbb{R}^{1\times N}$, and $\hat{\vec{f}}^i_t$ together with the fNCC feature $\hat{\vec{q}}^i_t$ is concatenated and used as the input to the separation module. The MISO separation module generates $\vec{g}^1_t \in \mathbb{R}^{1\times N}$, a feature vector for the reference channel, and a context decoder receives the concatenation of $\vec{F}^i_t$ and $\vec{g}^1_t$ to decode the filters $[\hat{\vec{h}}^1_{t-C}, \ldots, \hat{\vec{h}}^1_t, \ldots, \hat{\vec{h}}^1_{t+C}]$ of the context, where the estimated filter $\hat{\vec{h}}^1_t$ has the same shape as $\vec{f}^1_t$. The filters are then applied to the encoder outputs, and mean-pooling is applied across the time dimension:
\begin{align}
    \vec{z}^1_t = \frac{1}{1+2C}\sum_{j=0}^{2C}\vec{f}^1_{t-C+j}\odot \hat{\vec{h}}^1_{t-C+j}
\label{eqn:mask}
\end{align}
where $\odot$ denotes the Hadamard product. The implicit ``filter-and-sum'' operation is thus defined on the context. Figure~\ref{fig:flowchart} (D) shows the context encoder and decoder.

\section{Experiment configurations}
\label{sec:config}
\begin{table*}[!ht]
	\scriptsize
	\centering
	\caption{Experiment results with various model configurations. SI-SDRi is reported on decibel scale.}
	\label{tab:adhoc}
	\begin{tabular}{c|c|c|cccc|c}
		\thline
		\multirow{2}{*}{\thead{Model}} & \multirow{2}{*}{\thead{\# of param.}} & \multirow{2}{*}{\thead{\# of mics}} & \multicolumn{4}{c|}{\thead{Overlap ratio}} & \multirow{2}{*}{\thead{Average}} \\
		\cline{4-7}
		& & & $<$25\% & 25-50\% & 50-75\% & $>$75\% \\
		\thline
		\multicolumn{1}{l|}{FaSNet} & 2.9M & \multirow{7}{*}{2 / 4 / 6} & 15.0 / 15.3 / 14.8 & 10.7 / 11.1 / 11.6 & 8.6 / 9.2 / 9.3 & 5.4 / 7.0 / 7.0 & 9.7 / 10.8 / 10.9 \\
		\cline{1-2}\cline{4-8}
		\multicolumn{1}{r|}{+MISO} & 2.9M & & 14.8 / 15.5 / 15.7 & 10.4 / 11.3 / 11.9 & 8.5 / 9.0 / 9.4 & 5.0 / 6.8 / 7.1 & 9.5 / 10.8 / 11.2 \\
		\multicolumn{1}{r|}{+fNCC} & 3.0M & & 14.5 / 14.8 / 14.4 & 10.1 / 11.0 / 11.3 & 8.3 / 8.9 / 9.0 & 4.9 / 6.7 / 6.9 & 9.3 / 10.4 / 10.6 \\
		\cline{1-2}\cline{4-8}
		\multicolumn{1}{r|}{+MISO+fNCC} & 2.9M & & 15.0 / 15.7 / 15.7 & 10.6 / 11.4 / 12.2 & 8.4 / 9.4 / 9.6 & 5.3 / 7.4 / 8.0 & 9.7 / 11.1 / 11.6 \\
		\multicolumn{1}{r|}{+MISO+implicit} & 2.9M & & 14.2 / 14.9 / 15.2 & 9.8 / 10.9 / 11.3 & 7.7 / 8.1 / 8.7 & 4.6 / 5.7 / 6.1 & 8.9 / 10.0 / 10.6 \\
		\cline{1-2}\cline{4-8}
		\multicolumn{1}{r|}{+MISO+implicit+fNCC} & 3.0M & & 15.3 / 16.0 / 16.1 & 10.9 / 11.8 / 12.5 & 8.5 / 9.6 / 10.1 & 5.7 / 7.6 / 8.3 & 9.9 / 11.4 / 12.0 \\
		\cline{1-2}\cline{4-8}
		\multicolumn{1}{r|}{+MISO+implicit+fNCC+context} & 3.0M & & \textbf{15.6} / \textbf{16.4} / \textbf{16.5} & \textbf{11.2} / \textbf{12.4} / \textbf{12.9} & \textbf{9.0} / \textbf{10.1} / \textbf{10.3} & \textbf{5.8} / \textbf{7.9} / \textbf{8.8} & \textbf{10.2} / \textbf{11.8} / \textbf{12.3} \\
		\thline
	\end{tabular}
\end{table*}

\subsection{Dataset}

We evaluate our approach on a simulated ad-hoc multi-channel two-speaker noisy speech dataset. 20000, 5000 and 3000 4-second long utterances are simulated for training, validation and test sets, respectively. For each utterance, two speech signals and one noise signal are randomly selected from the 100-hour Librispeech subset \cite{panayotov2015librispeech} and the 100 Nonspeech Corpus \cite{web100nonspeech}, respectively. The overlap ratio between the two speakers is uniformly sampled between 0\% and 100\%, and the two speech signals are shifted accordingly and rescaled to a random relative SNR between 0 and 5 dB. The relative SNR between the power of the sum of the two clean speech signals and the noise is randomly sampled between 10 and 20 dB. The transformed signals are then convolved with the room impulse responses simulated by the image method \cite{allen1979image} using the gpuRIR toolbox \cite{diaz2020gpurir}. The length and width of all the rooms are randomly sampled between 3 and 10 meters, and the height is randomly sampled between 2.5 and 4 meters. The reverberation time (T60) is randomly sampled between 0.1 and 0.5 seconds. After convolution, the echoic signals are summed to create the mixture for each microphone. The ad-hoc array dataset contains utterances with 2 to 6 microphones, where the number of utterances for each microphone configuration is set equal.

\subsection{Model configurations}

The original FaSNet with transform-average-concatenate (TAC) module \cite{luo2020end} is used as the backbone architecture as well as the baseline. The frame size $L$ and the context size $W$ are both set to 16~ms (256 points), and the overlap ratio between frames is set to 50\%. In iFaSNet, we use the same configuration of $L$ and $W$ (with $C=2$ as discussed in Section~\ref{sec:fasnet}). The architecture for the context encoding and decoding modules has two choices:
\begin{enumerate}
    \item \textit{MLP}: $\vec{F}^i_t$ is flattened to a vector with shape $1\times N(1+2C)$, and a multi-layer perceptron (MLP) is used to map the vector into $\hat{\vec{f}}^i_t$. Another MLP is used to map the frame-level concatenation of $\vec{g}^1_t$, the output of the MISO separation module, and each feature in $\vec{F}^i_t$ to estimate the filters for the entire context.
    \item \textit{RNN}: $\vec{F}^i_t$ is passed to an RNN layer followed by a mean-pooling operation across time to generate $\hat{\vec{f}}^i_t$. The output of the MISO separation module $\vec{g}^1_t$ is concatenated to each feature in $\vec{F}^i_t$ and passed to another RNN layer to decoder the filters.
\end{enumerate}
We use the RNN configuration as empirically it leads to better performance than the MLP configuration with a same model size. We use two BLSTM layers for the context encoding and decoding modules, respectively. The implementations of both the original FaSNet and the iFaSNet are available online\footnote{\url{https://github.com/yluo42/TAC}}.

\subsection{Training configurations}

All models are trained for 100 epochs with the Adam optimizer \cite{kingma2014adam} with an initial learning rate of 0.001. Signal-to-noise ratio (SNR) is used as the training objective for all models. The learning rate is decayed by 0.98 for every two epochs. Gradient clipping by a maximum gradient norm of 5 is always applied for proper convergence of DPRNN-based models. Early stopping is applied when no best validation model is found for 10 consecutive epochs. No other training tricks or regularization techniques are used. Auxiliary autoencoding training (A2T) is applied to enhance the robustness on this reverberant separation task \cite{luo2020distortion}. 

\section{Results and discussions}
\label{sec:result}
Table~\ref{tab:adhoc} presents the experiment results of the baseline FaSNet and various configurations of iFaSNet. We first note that the results for FaSNet is lower than the results reported in \cite{luo2020end}, and this is due to the use of A2T during training. We suspect that this is mainly because the filter-and-sum problem formulation makes A2T training harder and affects the separation performance. We do not dive deeper into the role of filter-and-sum operation in A2T as it is beyond the scope of this paper. For ablation experiments, we explore the effect of introducing MISO configuration or replacing the original tNCC to fNCC. We observe that modifying the MIMO configuration into MISO configuration has no harm on the overall performance, while using fNCC instead of tNCC in the MIMO setting leads to worse performance. A possible explanation for this is that fNCC cannot explicitly capture cross-channel delay information, which is important for the MIMO time-domain filter-and-sum operation. 

The second step for the ablation experiments starts with setting the MISO configuration as default. Unlike the MIMO+fNCC configuration, MISO+fNCC configuration is able to improve the performance especially when the number of available channels is large. Comparing with the MISO+tNCC configuration, the improvement is more significant at higher overlap ratios. However, applying implicit filtering together with the tNCC feature leads to a drastic performance degrade. Since implicit filtering without context is equivalent to the ``masking'' configuration in various existing multi-channel end-to-end systems \cite{gu2019end, gu2020temporal, gu2020enhancing}, the result implies that the cross-channel feature needs to be carefully selected to align with the formulation of separation.

The third ablation experiment combines MISO, implicit filtering and fNCC feature and achieves better performance than any previous configurations. Note that comparing with the MISO+implicit configuration, such result indicates that the fNCC feature is a suitable cross-channel feature for the implicit filtering approach. The ablation experiments end with applying context-aware filtering into the system and result in our final design for iFaSNet. Note that the context size is set to $C=2$ as mentioned in Section~\ref{sec:context}, and the effect of other context sizes and window sizes is left for future work.

\section{Conclusion}
\label{sec:conclusion}
In this paper, we explored ways to improve a previously proposed end-to-end multi-channel speech separation system, the filter-and-sum network (FaSNet). We considered four different aspects to modify the original model: the multi-input-single-output (MISO) problem formulation, the implicit filtering in latent spae, the feature-level normalized cross correlation feature for cross-channel information, and the context-aware filtering operation. We named our modification to FaSNet as the implicit FaSNet (iFaSNet). Ablation experiment results showed that the iFaSNet combining such four modifications can lead to a significant performance improvement across various configurations.

\section{Acknowledgments}
This work was funded by a grant from the National Institute of Health, NIDCD, DC014279; a National Science Foundation CAREER Award; and the Pew Charitable Trusts.

\bibliographystyle{IEEEtran}
\bibliography{refs}

\end{document}